\documentclass[12pt]{iopart}
\usepackage{graphicx}
\usepackage{color}
\usepackage{iopams}

\usepackage[normalem]{ulem}
\begin{document}
 
 \title[]{Finite-size effects in the dynamics of few bosons in a ring potential}

\author{G.~Eriksson$^1$, J.~Bengtsson$^1$, E. \" O.~Karabulut$^{2}$, G.~M.~Kavoulakis$^3$, and S.~M.~Reimann$^1$}
\address{$^1$Mathematical Physics and NanoLund, LTH, Lund University, P. O. Box 118, SE-22100 Lund, Sweden}
\address {$^2$Department of Physics, Faculty of Science, Selcuk University, TR-42075 Konya, Turkey}
\address{$^3$Technological Education Institute of Crete, P. O. Box 1939, GR-71004, Heraklion, Greece}

\date{\today}

\begin{abstract}
We study the temporal evolution of a small number $N$ of ultra-cold bosonic atoms confined 
in a ring potential. Assuming that initially the system is in a solitary-wave solution of the corresponding mean-field 
problem, we identify significant differences in the time evolution of the density distribution of the atoms when it 
instead is evaluated with the many-body Schr\"odinger equation. Three characteristic timescales are derived: the first 
is the period of rotation of the wave around the ring, the second is associated with a ``decay" of the density variation, 
and the third is associated with periodic ``collapses" and ``revivals" of the density variations, with a factor of 
$\sqrt N$ separating each of them. The last two timescales tend to infinity in the appropriate limit of large $N$, in 
agreement with the mean-field approximation. These findings are based on the assumption of the initial state being a mean-field state. We confirm this behavior by comparison to the exact solutions for a few-body system stirred by an external potential. We find that the exact solutions of the driven system exhibit similar dynamical features.
\end{abstract}

 \maketitle

\section{Introduction}\label{Sec1}

The field of cold atomic gases has been progressing in various ways in recent years. Experimentally, it has become 
possible to confine and rotate Bose-Einstein condensed atoms in traps with a topology different from the usual harmonic 
confinement, such as for example in annular and toroidal traps~\cite{Gupta2005, Olson2007, Ryu2007, Sherlock2011, 
Ramanathan2011, Moulder2012, Beattie2013, Ryu2014, Eckel2014, Navez2016}. Another relatively new direction is towards 
systems with a reduced number of atoms $N$, all the way down to the few-body regime, see e.g. Ref.~\cite{Wenz2013}. 
In this regime, the physical properties of the system may deviate significantly from those predicted by the mean-field Gross-Pitaevskii equation,
  which relies on the assumption of a single product many-body state. Although the mean-field approximation is very 
successful in describing dilute bosonic systems in the large-$N$ limit, it is still an open issue to what extent 
it is applicable to smaller systems. In Refs.~\cite{Jackson2001, Cremon2013, 
Cremon2015, Roussou2015} a comparison is made between the stationary solutions obtained within the mean-field
approximation and those of the full many-body problem, to unravel the finite-$N$ differences between the two approaches. 

Here, we consider the temporal evolution of $N$ bosonic atoms rotating in a ring-shaped confinement. For such systems, the time-dependent Gross-Pitaevskii equation supports solitary wave solutions~\cite{Scott1973,Rajaraman1987,Zakharov1973}. Their time evolution is trivial, since the corresponding single-particle density distribution propagates around the ring without any change of shape. Much theoretical work concerns solitary-wave states beyond the mean-field description~\cite{Wadati1984,Wadati1985,Lai1989,Girardeau2000,Dziarmaga2002,Berman2004,Li2005,Kanamoto2010,Zollner2011,Sato2012,Heimsoth2012,Delande2014,Syrwid2015,Syrwid2016,Roussou2017}. However, to the best of our knowledge, a direct comparison to the full many-body dynamics in the few-body limit has not yet been performed.
Contrary to the Gross-Pitaevskii approach, 
the time-dependent many-body Schr\"odinger equation is generally not
expected to support solitary-wave solutions for finite $N$.   On the
other hand, in the large-$N$ limit we do expect the mean-field
description to be valid and solitary-wave states to form. This calls
for a systematic study of the finite-$N$ effects in the temporal
evolution of the solitary-wave states predicted by mean-field theory. In other words, we choose to start in a single product many-body state derived from the Gross-Pitaevskii equation, but time-evolve it using the full many-body Hamiltonian $\hat{H}$. As later shown, an almost identical temporal evolution can, however, also be found in the exact solutions for an initially stirred few-body system, without invoking the mean-field approximation.
We restrict our study to strictly one-dimensional systems
with periodic boundary conditions. 
For simplicity we choose to sample the single-particle density
distribution $\rho$ in time, at a fixed location in space. 
From $\rho $ we observe three different time scales: 
(i) the time scale $T_{GP}$ for a single revolution of 
the solitary wave around the ring, 
(ii) the time scale $\tau _s$ associated with the collapse of the initial solitary wave state
and (iii)  the time scale $T_A$ 
of the periodic reappearance of the solitary-wave  state.

The plan of the paper is as follows. In Sec.~\ref{Sec2}, we briefly
discuss the general form of the many-body Hamiltonian, along with the
corresponding Gross-Pitaevskii equation. 
In Sec.~\ref{Sec3} we consider weak interactions.  In this case, an 
approximate description of the dynamics based on a two-state model is
discussed, where the time-evolution is obtained analytically. Within
this model, we show that the dynamics generated by a full 
many-body Hamiltonian approaches that of the mean-field solution in the 
appropriate limit of large $N$. 
The case of stronger interactions, where we need to go
beyond the two-state model, is treated numerically in Sec.~\ref{Sec4}. 
In Sec.~\ref{Sec5}, we study the full many-body solutions for an explicitly driven system, and give a summary of our results and some general 
conclusions in Sec.~\ref{Sec7}.

{\section{Model} \label{Sec2}}

Let us now consider $N$ repulsive bosons in a one-dimensional confinement
with periodic boundary conditions. The inter-atomic interactions, here
assumed to be elastic s-wave atom-atom collisions, are modeled by the
pseudo-potential $g\delta(\theta_i-\theta_j)$, where $g>0$ is the
interaction strength and $0 \leq \theta_i \leq 2\pi$ the angular
coordinate of the $i$th particle. In an actual toroidal, or annular
potential, the one-dimensional treatment of the system is valid
provided that the interaction energy is much smaller than the energy
required to excite the system in the transverse direction, in which
case the corresponding degrees of freedom are frozen.

If ${\hat a}_m$ and ${\hat a}_m^{\dagger}$ are the annihilation and
creation operators of an atom in the single-particle state $\phi_m (\theta) =
e^{i m \theta}/\sqrt{2 \pi}$ with angular momentum $m \hbar$, the
Hamiltonian of the system reads 
\begin{eqnarray}
  {\hat H} = \epsilon \sum_m m^2 {\hat a}_m^{\dagger} {\hat a}_m 
  +  \frac{1}{2} { \frac{ g}{ 2\pi}} \sum_{m+n=k+l} {\hat a}_m^{\dagger} {\hat a}_n^{\dagger} a_k a_l,
\label{ham11}
\end{eqnarray}
where $\epsilon= {\hbar^2} /{(2 M R^2)}$ is the kinetic energy per
particle, with $M$ being the atom mass and $R$ being the radius of the
ring. There are thus two energy scales in the problem:
the kinetic energy $\epsilon$, associated with the motion
of the atoms along the ring, and the interaction energy per particle,
which for a homogeneous gas is equal to $g(N-1)/ (4\pi)$. From these
energies, we introduce the dimensionless quantity
\begin{eqnarray}
\gamma = \frac{g(N-1)}{2\pi\epsilon}~. \label{gamma}
\end{eqnarray}
The corresponding time-dependent mean-field Gross-Pitaevskii
equation for the order parameter $\psi(\theta, t)$ reads
\begin{eqnarray}
  i { \hbar}\frac {\partial \psi(\theta, t)} {\partial t} = - {\epsilon}\frac {\partial^2 \psi(\theta, t)} {\partial \theta^2} 
  + {g(N-1)}|\psi(\theta,t)|^2 \psi(\theta,t).\label{TDGP}
\end{eqnarray}
The solitary-wave solutions of Eq.~(\ref{TDGP}) are of the form
$\psi(\theta, t) = \psi(\theta- \Omega_{GP} t)$, where $\Omega_{GP}$ is the
angular frequency of rotation of the wave, and satisfy
\begin{eqnarray}
  - i { \hbar}\Omega_{GP} \frac {\partial \psi(z)} {\partial z} = - {\epsilon}\frac {\partial^2 \psi (z)} {\partial z^2} 
  + { g(N-1)} |\psi(z)|^2 \psi(z), \label{TDGPsol}
\end{eqnarray}
where $z = \theta - \Omega_{GP} t$. For periodic boundary conditions, the
solutions of Eq.~(\ref{TDGPsol}) are Jacobi elliptic functions
\cite{Carr2000,Smyrnakis2010}. Equivalently one may view the above as
an ``yrast" problem in the mean-field description, namely the 
minimization of the energy for a fixed value of the angular momentum,
where $\Omega _{GP}$ is a Lagrange multiplier \cite{Jackson2011}.
 
\section{Two-state model}\label{Sec3}

For weak interactions ($ \gamma  \ll 1$) and 
for total angular momenta $0 \leq L \leq N\hbar$, it is
sufficient to consider the contribution from the single-particle
states $\phi_0(\theta)$ and $\phi_1(\theta)$ alone. Introducing the (two-state) trial
function
\begin{eqnarray}
 \psi(\theta, t) = c_0(t) \phi_0(\theta) + c_1(t) \phi_1(\theta)
\end{eqnarray}
in the time-dependent Gross-Pitaevskii equation~(\ref{TDGP})
produces the following set of differential equations:
\begin{eqnarray} 
  i { \hbar}\frac{\partial}{\partial t}c_0(t) &=& { \epsilon}\gamma (1 + |c_1(t)|^2) c_0(t), \label{TDGPc0}
  \\
  i { \hbar}\frac{\partial}{\partial t}c_1(t) &=& { \epsilon}[1 + \gamma (1 + |c_0(t)|^2)] c_1(t).\label{TDGPc1}
\end{eqnarray} 
The solution of
Eqs.~(\ref{TDGPc0}) and~(\ref{TDGPc1}), with the phase-convention of 
real-valued coefficients at $t=0$, is in turn given by
\begin{eqnarray}
  c_0(t) &=& \sqrt{1 - \ell}  e^{-i { \epsilon}\gamma (1 + \ell) t{ /\hbar}},
  \\
  c_1(t) &=& \sqrt{\ell}  e^{-i { \epsilon}[1+\gamma (2 - \ell)] t{ /\hbar}},
\end{eqnarray}
where $0 \le \ell \le  1$  and 
$\ell\hbar$ is the expectation value of the angular momentum per particle.
Other values of $\ell$ can
be treated in a similar fashion~\cite{Bloch1973}. Having established the
time evolution of the order parameter, the single-particle density 
$ \rho_{GP}(\theta,t)$ can be retrieved as
\begin{eqnarray}
 2\pi \rho_{GP}(\theta, t) &=& 2\pi|\psi(\theta,t)|^2 \nonumber \\
  &=& { 1+ 2\sqrt{\ell (1 - \ell)} \cos\left( \theta - \Omega_{GP} t \right)}, 
\label{densmf}
\end{eqnarray}
where 
\begin{eqnarray}
 \Omega_{GP}  = {\left[ 1 + \gamma (1 - 2 \ell)\right] \frac{\epsilon }{\hbar }}
\label{ommfl}
\end{eqnarray}
is the angular frequency of rotation of the solitary wave solution. Hence,
the density is given by a sinusoidal wave, with its center located
(arbitrarily) at $\theta=0$ when $t=0$. For a fixed value of $\theta $
we see a periodic modulation of $\rho_{GP}$ in time, with periodicity
\begin{eqnarray}
T_{GP} = \frac{2\pi}{\Omega_{GP}} = \frac{2\pi}{\left[1 + \gamma (1 - 2 \ell)\right]}\frac{\hbar }{\epsilon}. \label{TGP} 
\end{eqnarray}

Let us now turn to the dynamics of the corresponding $N$-body state
vector $|\Psi (t)\rangle $. For consistency, we work in the
restricted (many-body) Hilbert-space spanned by  
{ $|\phi_0\rangle$} and $| \phi_1\rangle$ alone. We may then write
\begin{eqnarray}
|\Psi(t)\rangle = \sum_{n=0}^{N} d_n(t)| N-n, n\rangle, \label{PsiExp}
\end{eqnarray}
where in this notation, 
$n$ is the occupation number of $|\phi_1\rangle$ (and $N-n$
that of $|\phi_0\rangle$). We further choose the initial state
vector, at $t=0$, to be the one given by the mean-field single-product state
associated with $\psi(\theta,0)$  discussed above. In other
words,
\begin{eqnarray}
| \Psi (0) \rangle &=& \frac 1 {\sqrt{N!}} \left(\sqrt{1-\ell} {\hat a}_0^{\dagger} + \sqrt{\ell} {\hat a}_1^{\dagger} \right)^N
 | 0 \rangle,
\end{eqnarray}
which implies that 
\begin{eqnarray}
d_n(0)=\sqrt{\frac {{N!}} {{(N-n)! n!}}}\left(1-\ell\right)^{(N-n)/2} \ell^{n/2}.\label{dtwostate}
\end{eqnarray}
With identical initial states, we may now compare the 
properties of the system given by $| \Psi(t)\rangle$ 
with those of the order parameter $\psi(\theta,t)$ at later times.  
Here, the time-evolution of the many-body state $| \Psi(t)\rangle$  is given by the
time-dependent Schr\"{o}dinger equation,
\begin{eqnarray} 
i\hbar
\frac {\partial} {\partial t} |\Psi(t) \rangle = {\hat H} |\Psi(t)\rangle.
\end{eqnarray}
The fact that the total angular momentum $N\ell\hbar$ is conserved
renders the time-evolution of $| \Psi(t)\rangle $ trivial
within the two-state model,
\begin{eqnarray}
d_n(t) = d_n(0)e^{-i{\cal E}_nt/{\hbar}},
\end{eqnarray}
with 
\begin{eqnarray}
 {\cal E}_n = n{\epsilon}+  {g} \left[ N (N-1) + 2n (N-n)\right] /(4\pi) .
\label{disrel}
\end{eqnarray}
Finally, the
single-particle density $\rho(\theta,t)$ associated with $| \Psi(t)\rangle $ is
\begin{eqnarray}
2\pi \rho(\theta,t) &=& \frac{2\pi }{N} \sum_{i,j=0}^1\phi^*_j(\theta) \phi_i(\theta) \langle \Psi(t) |   {\hat a}^\dagger_j {\hat a}_i  | \Psi(t) \rangle \nonumber \\
& = &  1 + 2 \sqrt{\ell(1-\ell)} A(t) \cos\left[\theta - \Omega(t)t \right], \label{dens}
\end{eqnarray}
with
\begin{eqnarray}
  &&A(t) = \left[1 - 4 \ell (1 - \ell) \sin^2\left({\frac{gt}{2\pi \hbar}}\right)\right]^{{(N-1)}/{2}}, \label{amplit}\\
&&\Omega(t) = {\frac{\epsilon}{\hbar}}  + \frac{(N-1) \omega(t)}{t} ,\\
&&\tan \left[\omega(t)\right] =( 1 - 2 \ell) \tan\left({\frac{gt}{2\pi \hbar}}\right).
 \label{phase}
\end{eqnarray} 
The density $\rho (\theta,t) $ in Eq.~(\ref{dens}) should be compared with the
one in Eq.~(\ref{densmf}). Contrary to $\rho_{GP}(\theta,t)$, the above
expression for $\rho (\theta,t) $ includes a time-dependent rotation frequency $\Omega(t)$, as well
as an additional time-dependent amplitude modulation, $A(t)$.

For times $t \ll 2\pi\hbar/ g  = (N-1)\hbar/(\gamma \epsilon)$, we can expand both the left and right side of Eq.~(\ref{phase}),
\begin{eqnarray}
\omega(t) \approx  {\frac{(1 - 2 \ell) gt}{2\pi \hbar}},
\end{eqnarray}
and approximate the angular frequency of rotation with
\begin{eqnarray}
\Omega(t) \approx \frac{\epsilon}{\hbar} + \frac{(N-1)(1 - 2 \ell) g}{2\pi \hbar} = \left [1 + \gamma (1 - 2 \ell)\right]  {\epsilon \over \hbar }, \label{omtomfl}
\end{eqnarray}
which is identical to the frequency $\Omega_{GP}$ obtained for the
solitary wave in the mean-field approach, see Eq.~(\ref{ommfl}). Note that by
decreasing the interaction strength $g$ we extend the
time-period during which the approximation $\Omega(t) \approx
\Omega_{GP}$ is valid. Similarly, in the limit $N\to\infty$ and
for fixed values of $\gamma$  and $\epsilon$, we find that
$\Omega(t)\to\Omega_{GP}$. We also observe that $\Omega_{GP}$ agrees,
to  leading order in $N$, with the derivative of the dispersion
relation, Eq.~(\ref{disrel}), 
\begin{eqnarray}
\left. \frac{d{\cal E}_n}{d (n\hbar )}\right|_{n=N\ell } = 
\biggl [1+\gamma (1-2\ell ) \frac{N}{N-1}\biggr ] 
\frac{\epsilon}{\hbar}\approx \Omega_{GP}, 
\label{OmegaFromDisp}
\end{eqnarray}
 where we recall that $L=n\hbar$ in the two-state 
model.   
The amplitude modulation $A(t)$ in Eq.~(\ref{amplit}) has a periodicity of 
\begin{eqnarray}
T_A = \frac{2\pi^2\hbar}{g} = \frac{(N-1)\pi }{\gamma }{\hbar\over \epsilon }. \label{TA}
\end{eqnarray}
The corresponding periodic behavior of the wave function is expected. Similar collapses and revivals are seen for many quantum systems where restricted model spaces are adequate, see, e.g., Ref.~\cite{Eberly1980}. Here, the analytic form of $T_A$ allows us to directly relate the periodicity of $A(t)$ to the interaction strength and number of particles in the considered system. Specifically,   
 for fixed values of $\gamma$ and $\epsilon$, $T_A$ grows
linearly with $N$.  
For times $t \ll 2\pi\hbar/ g $, i.e., when $\Omega \approx
\Omega_{GP}$, $A(t)$ alone determines how closely $|
\Psi(t)\rangle $ resembles the mean-field solitary-wave solution. In
particular, $\rho(\theta, t) \approx \rho_{GP}(\theta,t)$ when $A(t)
\approx 1 = A(t=0)$. To quantify the behavior of $| \Psi(t)
\rangle$, we use 
$| A(t) - A(0) | \leq A(0)/2$ as a criterion for a
solitary-wave state. The system thus exhibits solitary-wave 
behavior for times $0 \leq t \leq \tau_s$ (assuming that such a limit
exists), where 
\begin{eqnarray}
\tau_s &=& \frac{2\pi\hbar}{g}\arcsin\sqrt{\frac{1-2^{-2/(N-1)}}{4\ell(1-\ell)}} \nonumber \\
&=& \frac{\hbar(N-1)}{\gamma\epsilon} \arcsin\sqrt{\frac{1-2^{-2/(N-1)}}{4\ell(1-\ell)}}~.\label{taus}
\end{eqnarray}
For large $N$, we may use $1-2^{-2/(N-1)} \approx 2\ln2/(N-1)$ and
write
\begin{eqnarray}
\tau_s &\approx&  \frac{\hbar}{\gamma\epsilon} \sqrt{\frac{(N-1)\ln2}{2\ell(1-\ell)}}.
 \end{eqnarray}
Therefore, for large $N$, there is generally a clear hierarchy of
timescales,
\begin{eqnarray}
T_{GP} \ll \tau_s \ll T_{A},
\end{eqnarray} 
with a factor of $\sqrt N$ separating each of them (for some fixed
$\gamma$ and $\epsilon $). 
When time-evolving a mean-field solitary-wave state
using the full many-body Hamiltonian, the system will initially
(for $t < \tau_s$) mimic the behavior of a solitary wave with 
periodicity $T_{GP}$. Unlike the mean-field time-evolution, a more
homogenous density distribution is subsequently observed (with $A(t)\approx
\left[ 1 - 4 \ell \left(1 - \ell\right)\right]^{(N-1)/2}$). However, at even later times $T_A$, the solitary-wave behavior reappears.

\begin{figure}
\center
\includegraphics[width=0.75\linewidth]{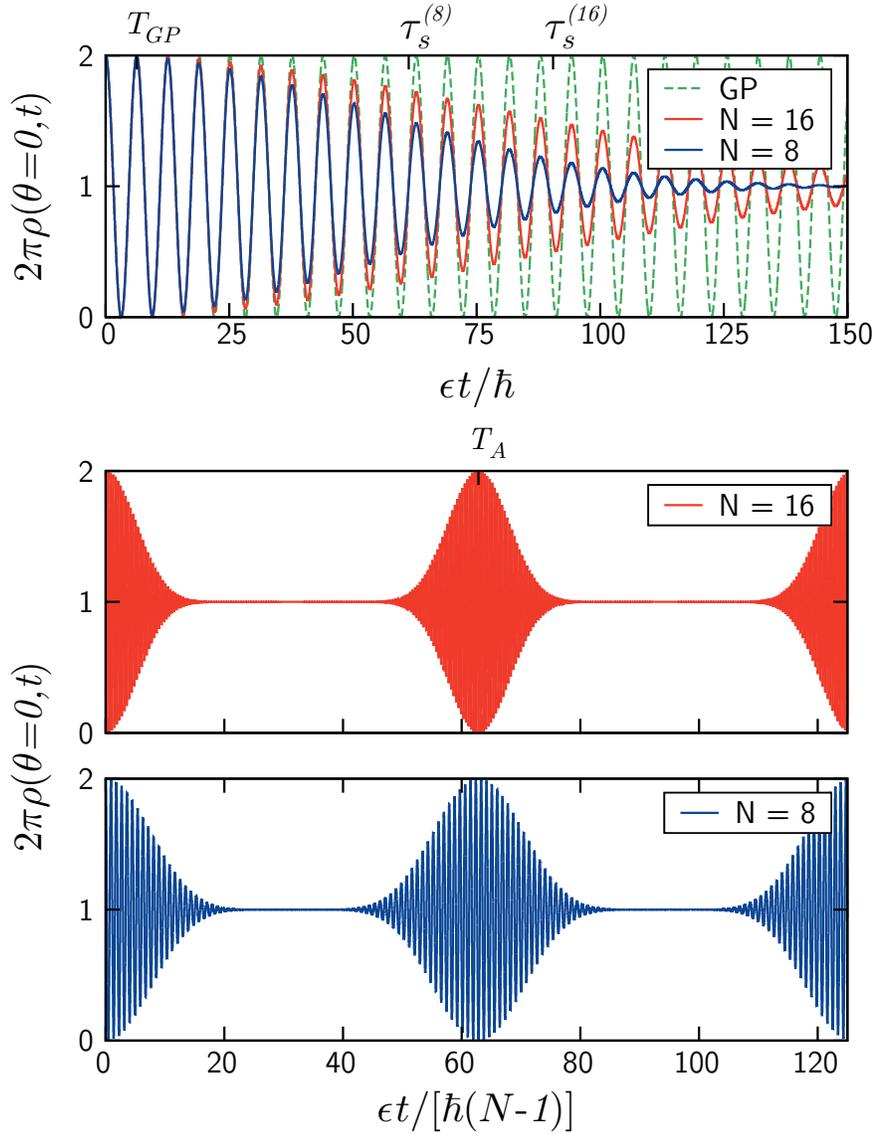}
\caption{(Color online) Single-particle density distribution
  {(within the two-state model)} evaluated at $\theta =0$ as
  a function of time, for $\gamma = 0.05$ and $\ell = 1/2$.  The top
  panel shows the collapse of the initial solitary-wave state when
  propagated with the full Hamiltonian for $N=8$ (blue solid line) and
  $N=16$ (red solid line).  The corresponding mean-field solution is
  shown for comparison (green dashed line).  Note the different time
  scales $T_{GP}$ and $\tau _s^{(N)}$, where the latter depends on
  $N$.  The two lower panels show the periodic revival with time
  $T_A$ of the solitary-wave behavior, for $N=8$ and $N=16$. Note
  here that the unit of time is scaled with a factor $(N-1)$. }
\label{fig1}
\end{figure}

In Fig.~\ref{fig1} we compare the two expressions for the
single-particle density obtained within the two-state model, i.e., $\rho (\theta,t)$
and $\rho_{GP}{(\theta,t)}$ in Eqs.~(\ref{densmf}) and (\ref{dens}), 
respectively. More specifically, we evaluate the density of the
``dark" solitary wave, obtained for  $\ell = 1/2$, at $\theta = 0$ as
a function of time for weak repulsive interactions, $\gamma=0.05$, 
with $N=8$ and $N=16$. 
Increasing $N$ extends the time-period $\tau_s$ in which
$\rho{(0,t)}$ and $\rho_{GP}{(0,t)}$ approximately agree, as shown in the top
panel of Fig.~\ref{fig1}. In the two lower panels, we see the decay and revival of the solitary wave at a time-period $T_A$, which increases linearly
with $N$.

\bigskip

\section{Beyond the two-state model}
\label{Sec4}

We now turn to systems with stronger interactions and investigate
whether the characteristic dynamical features of the solitary-wave
states, identified in the weak-interacting limit, persist.

With an increase of $\gamma$ follows an increase in the number of
single-particles states $\phi_m{(\theta)}$ with non-negligible contributions to
the order parameter $\psi{(\theta,t)} $. In other words, for an adequate
description of the system, we need to go beyond the two-state model
discussed in Sec.~\ref{Sec3}. The necessary extension to a larger
basis means, in turn, that the initial state and its time-evolution
have to be evaluated numerically. Here, for example, we use an
exponential propagator in the Krylov subspace~\cite{Park1986} to solve
the time-dependent many-body Schr\"{o}dinger equation, and an
exponential Lawson scheme~\cite{Lawson1967} for the time-dependent
Gross-Pitaevskii equation.  The occupancy of the corresponding single-particle states 
decays rapidly with increasing values of $| m-\ell| $. Even for relatively
strong interactions surprisingly few single-particle states thus need
to be considered.  On the other hand,
based on Eq.~(\ref{taus}), we expect that with increasing
$\gamma $ also $N$ has to increase  if we want to keep
reasonable values of  $\tau _s \gtrsim T_{GP}$. The
computational workload of the full many-body problem grows rapidly
with $N$, setting a limit to the interaction strengths we may consider
numerically.

\begin{figure}
\center
\includegraphics[width=0.75\linewidth]{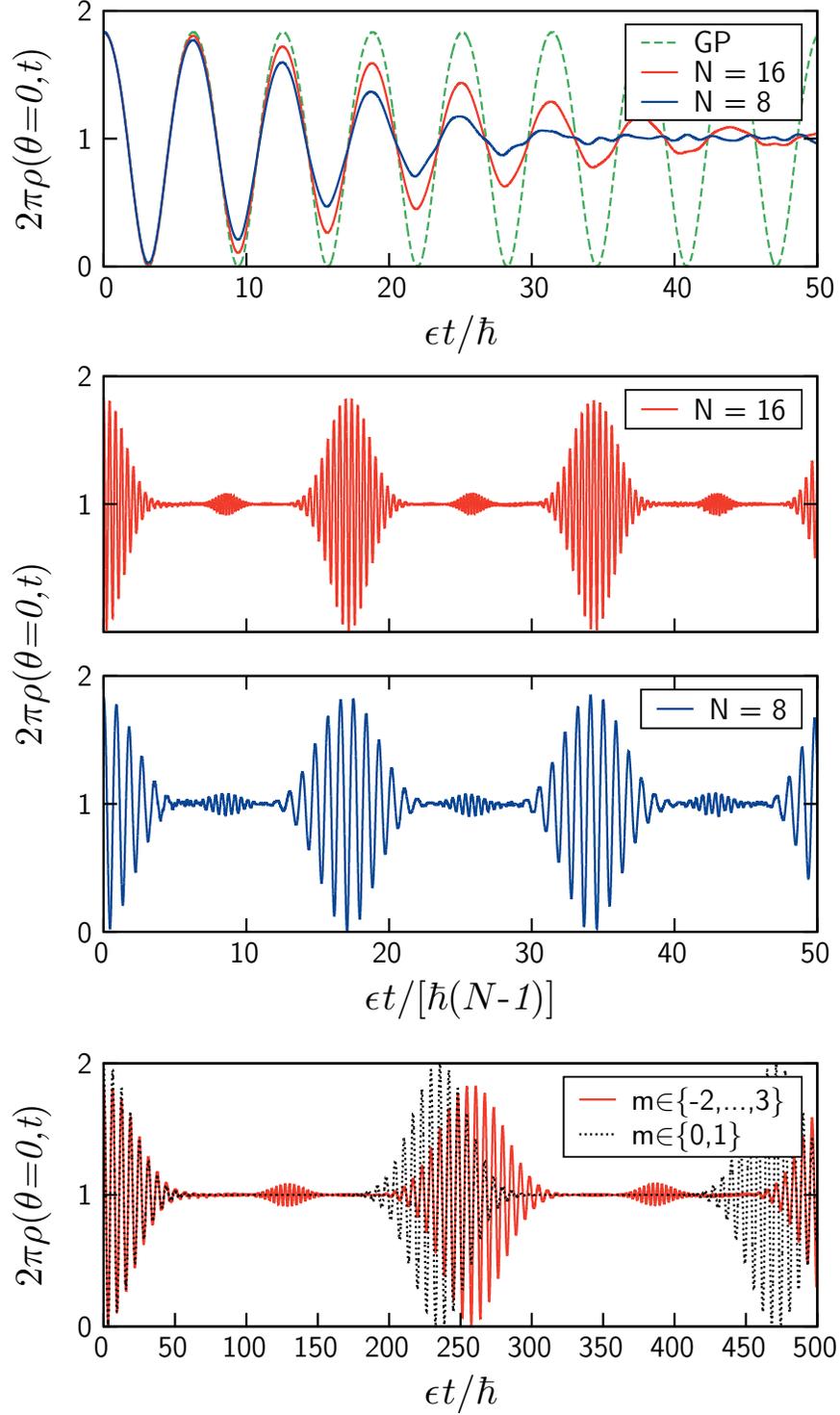}
\caption{{\it (Color online)} The top and middle panels are as in Fig.~\ref{fig1}, but for $\gamma = 0.2$, and single-particle states $m=-2,-1,0,1,2,3$. 
The lowest panel (here for $N=16$) shows that in this case, the two-state model (black dotted line) can not reproduce the actual periodicity $T_A$ of the system.}
\label{fig2}
\end{figure}

In Fig.~\ref{fig2}, we show (as in Fig.~\ref{fig1}) the
single-particle density at $\theta=0$ for the same case with $\ell =
1/2$, $N=8$ and $N=16$. However, we now consider a stronger
interaction, $\gamma = 0.2$, and include the single-particle states
with $m=-2,-1,0,1,2$ and $3$ for an adequate description. This truncation gives, in the case of $N=16$, a many-body basis of 20349 states. We observe
that the characteristic features discussed in the limit of weak
interactions remain. In the top panel, we see that the early time evolution
based on the full many-body Hamiltonian approaches that of its
mean-field equivalent in the limit of large $N$.  
As seen in the two middle panels, the dynamical
structure of $\rho$ for $\gamma=0.2$ looks very similar to what was
obtained for $\gamma=0.05$ (shown in Fig.~\ref{fig1}). Note, however,
that the maximum value is slightly lower and that there are some minor additional oscillations.  The periodicity $T_A$ of
the collapse and revival associated with the solitary-wave state still
seems to increase linearly with $N$, in a similar way as predicted by
Eq.~(\ref{TA}) for weak interactions. However, the fact that $T_A$
scales linearly with $N$, as in the two-state model, does {\it not}
imply that the restricted model can be used to extract the
actual periodicity of the system.  In the lowest panel of
Fig.~\ref{fig2}, we clearly see the different $T_A$ obtained
in a calculation limited to the states with $m=0$ and $m=1$ compared to that of the extended space.

\begin{figure}
\center
\includegraphics[width=0.75\linewidth]{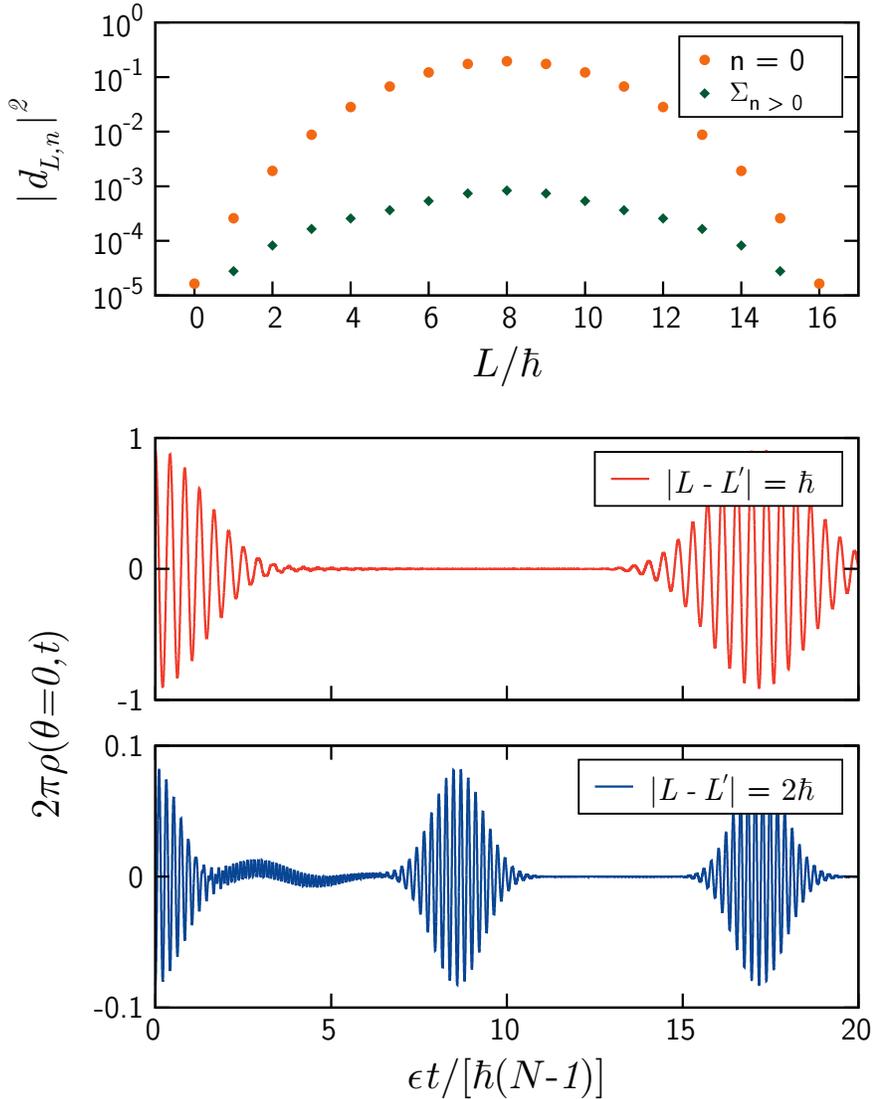}
\caption{{\it (Color online)} The top panel shows the population $| d_{L,n}|^2$ of the stationary solutions $|\Phi_{L,n}\rangle$, with total angular momentum $L$, for both $n = 0$ {\it (orange dots)} and the sum of all other states {\it (green diamonds)}. The middle and lower panels show the contributions to the single-particle density originating from the different values of $| L-L^{\prime}| = \hbar$ and $2\hbar$ respectively. In all panels we consider the case of $N=16$, $\gamma = 0.2$, $\ell=1/2$ and $m=-2,-1,0,1,2$ and $3$.}
\label{fig3}
\end{figure}

Finally, we examine the origin of the minor oscillations in $\rho(0,t)$ seen, e.g., close to $\epsilon t/[\hbar(N-1)]\approx 9$ in Fig.~\ref{fig2}, which are in sharp contrast to the homogeneous density distribution predicted by the corresponding two-state model.
  We expand the many-body state $|\Psi(t)\rangle$ in the many-body eigenstates $|\Phi_{L,n}\rangle$ of $\hat{H}$,
\begin{equation}
  |\Psi(t)\rangle = \sum_{L,n}d_{L,n}e^{-iE_{L,n}t/\hbar}|\Phi_{L,n}\rangle, \label{Psidecomposed}
\end{equation}
where $n=0,1,\ldots$  orders the different states with the same total angular momentum $L$ by their increasing eigenenergies $E_{L,n}$, and where $d_{L,n}$ are the (time-independent) expansion coefficients. With the expansion of $|\Psi(t)\rangle$ in Eq.~(\ref{Psidecomposed}), we may write the single-particle density as  
\begin{eqnarray}
2\pi \rho(\theta,t) &=& \frac{2\pi }{N}\sum_{L,L'}\sum_{n,n'}d^*_{L',n'}d_{L,n}e^{-i(E_{L,n}-E_{L',n'})t/\hbar} \times \nonumber \\ 
& & \times\sum_{i,j=m_{min}}^{m_{max}}\phi^*_j(\theta) \phi_i(\theta) \langle\Phi_{L',n'}|   {\hat a}^\dagger_j {\hat a}_i  |\Phi_{L,n}\rangle, \label{decomposed_dens}
\end{eqnarray}
where $m_{min}$ and $m_{max}$ specify the set $m=m_{min},\ldots,m_{max}$ of considered single-particle states $| \phi_m\rangle$. This numerical restriction in $m$ implies, in turn, that only combinations where $| L-L^{\prime}| \leq (m_{max}-m_{min}) \hbar$ contribute to $\rho$ in Eq.~(\ref{decomposed_dens}). As in Fig.~\ref{fig2}, we chose $m_{min}=-2$, $m_{max}=3$ and consider the case of $N=16$, $\gamma = 0.2$ and $\ell =1/2$. Now, in the top panel of Fig.~\ref{fig3}, we first show the population $| d_{L,0}|^2$ of the yrast state, $|\Phi_{L,0}\rangle$,  for each $L$, together with the combined population of the remaining (excited) eigenstates. Clearly, the populations of the low energetic yrast states dominate, with a population peak at $L=N\ell\hbar=8\hbar$. In fact, for such a weak interaction $\gamma$, the populations of the different yrast states are close to Gaussian shaped in $L$, resembling the distribution  of the corresponding two-state model, see Eq.~(\ref{dtwostate}).

Next, we turn to the decomposed contributions to $\rho(0,t)$ in Eq.~(\ref{decomposed_dens}), originating from different values of $| i - j| = | L-L^{\prime}|/\hbar$ in the summation over single-particle states. Due to the large influence of the yrast states, an (almost) static contribution $2\pi\rho(0,t)=1$ is obtained with $L=L^\prime$. The more interesting cases of $|L-L^\prime| = \hbar$ and $|L-L^\prime| = 2\hbar$ are shown in the lower two panels of Fig.~\ref{fig3}. As in the two-state model, which is limited to $|L-L^\prime| \leq \hbar$ and has a single many-body eigenstate for each $L$, the main dynamical features of $\rho(0,t)$ are captured by the interference of eigenstates with $|L-L^\prime| = \hbar$, producing collapses and revivals of the solitary-wave state. The minor additional oscillations in $\rho(0,t)$, seen at $\epsilon t/[\hbar(N-1)]\approx 9$ in Fig.~\ref{fig2}, originate primarily from the $|L-L^\prime| = 2\hbar$ contribution (see lowest panel of Fig.~\ref{fig3}). Intriguingly, the dynamical features in $\rho(0,t)$ originating from terms with $|L-L^\prime| = 2\hbar$ show striking similarities to that obtained with $|L-L^\prime| = \hbar$, although with a smaller amplitude as well as with a reduced periodicity and revival time of the additional oscillations. The reduced amplitude in the new oscillations, compared to the amplitude obtained with $|L-L^\prime|=\hbar$, may largely be explained by the smaller magnitudes of $\langle\Phi_{L',0}|   {\hat a}^\dagger_j {\hat a}_i  |\Phi_{L,0}\rangle$ when $|i-j|=2$. For the weak interaction strengths considered, the yrast states are largely dictated by the occupancies of the single-particle states $|\phi_0\rangle$ and $|\phi_1\rangle$. The comparable dynamical features, with collapses and revivals, seen in the two lower panels of Fig.~3 may be understood from the fact that $d^*_{L^\prime,0}d_{L,0}$ has a similar structure for $|L-L^\prime| = \hbar$ and for $|L-L^\prime| = 2\hbar$.
 Furthermore, $E_{L,0}$ is largely linear in $L$ (see the two-state model equivalent in Eq.~(\ref{disrel})) and, consequently, $E_{L+2,0}-E_{L,0} \approx 2(E_{L+1,0}-E_{L,0})$. The time-scales of $\rho(0,t)$ associated with $|L-L^\prime|=2\hbar$ are thus approximately half of those originating from $|L-L^\prime|=\hbar$. In other words, the periodicity and revival time of the additional solitary-wave state behavior, shown in the lowest panel of Fig.~\ref{fig3}, are roughly $T_{GP}/2$ and $T_A/2$ respectively. 
Increasing $|L-L^\prime|$ further, we find more contributions to $\rho(0,t)$ with even smaller amplitudes and shorter time-scales (not shown here). Eventually, for small enough amplitude modulations, also the more complex temporal behavior caused by the interference between yrast states and excited states becomes important.  In fact, the peculiar behavior of the   $|L-L^\prime|=2\hbar$ contribution to $\rho(0,t)$, seen between the first collapse and revival of the additional (minor) solitary-wave state, is caused by such interference terms.

\bigskip

\section{Dynamical stirring}
\label{Sec5}
To create a solitary-wave state experimentally, there are three main techniques that are used; phase-imprinting~\cite{Burger1999}, Laguerre-Gaussian 
beams~\cite{Andersen2006} and directly stirring the condensate with a potential barrier~\cite{Ramanathan2011}.

\begin{figure}
\center
\includegraphics[width=0.75\linewidth]{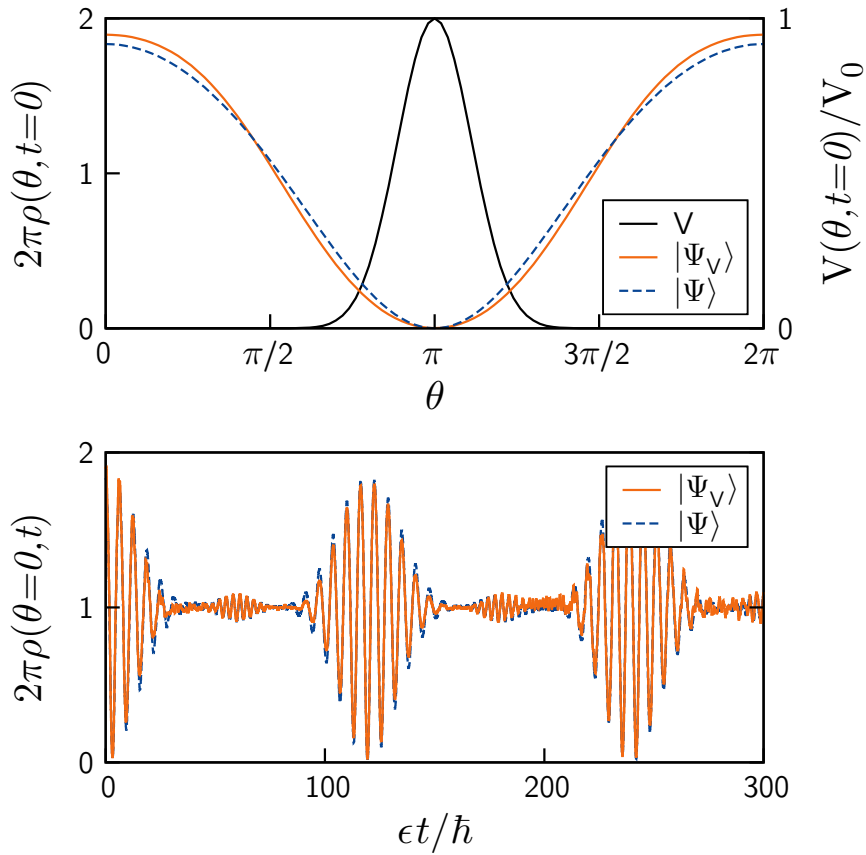}
\caption{{\it (Color online)} The top panel shows the stirring potential $V$ at $t=0$, where $q=32$, 
$\Omega _V = \hbar /(M R^2)$ and $V_0= 3.47 \epsilon $, and  the corresponding ground-state density {\it (orange solid line)} for $\gamma =0.2$ and $N=8$. Here, we consider  single-particle states with $m=-2,-1,0,1,2,3$. For comparison we also { include} the density of the mean-field product state  $| \Psi(0)\rangle$ {\it (blue dashed line)} obtained without $V$, that matches the energy and angular momentum of the {considered} many-body state  $| \Psi_V(0)\rangle$. The lower panel shows the free, non-driven time evolution, both for $| \Psi_V(t)\rangle$ (orange solid line) and $| \Psi(t)\rangle$ (blue dashed line), after the quench at $t=0$.  Also in this case, the two density distributions agree (the difference is hardly visible).}
\label{fig4}
\end{figure}

Here we examine the dynamic response when stirring  a few-body system.  We add to $\hat H$ a time-dependent external potential $V(\theta,t)$ that drives the system, 
\begin{equation}
V(\theta,t)=V_0 \sin^q {
\left(\frac{\theta-\Omega_Vt}{2}\right)
},
\end{equation}
{where $V_0>0$ is its amplitude}, $q$ is an integer,  and  $\Omega_V$ is the angular frequency of rotation. 
We further assume that the system remains in its ground state $| \Psi _V(t)\rangle $ in the rotating frame of reference. 
We now investigate the effect of this stirring potential on the density distribution.
In the top panel of Fig.~\ref{fig4} we show a particular choice of $V$ { at $t=0$} and the corresponding density distribution for the case of $\gamma =0.2$ and $N=8$.
Similar to the results shown in Fig.~\ref{fig1} and ~\ref{fig2}, the average angular momentum { per particle associated with $|\Psi _V(t)\rangle$ is $\hbar/2$}.   
Obviously,  the maximum of the potential $V(\theta,t)$ coincides with the minimum of $\rho (\theta,t)$, and vice versa. A similar density distribution is obtained { also in the absence of $V$} for the single product state $| \Psi{(t)} \rangle $ {with} $\langle \Psi{(t)}| \hat{H} |
\Psi{(t)}\rangle = \langle \Psi_V{(t)} | \hat{H} |
\Psi_V{(t)}\rangle$ and 
$\langle \Psi{(t)}| \hat{L} |
\Psi{(t)}\rangle = \langle \Psi_V{(t)}| \hat{L} |
\Psi_V{(t)}\rangle $.

By a sudden quench at $t=0$, we remove $V$ 
and examine the { new} 
 time-evolution { of $| \Psi_V(t)\rangle$} based on the time-independent original
Hamiltonian.  In the lower panel of Fig.~\ref{fig4} we 
observe collapses and revivals of the 
density modulation that appear similar to those of the 
solitary-wave state  described in Sec.~\ref{Sec3} and
Sec.~\ref{Sec4}. An almost identical behavior also follows from the full many-body time-evolution of the initial single product state { $| \Psi(0)\rangle$}, considered in the the top panel of Fig.~\ref{fig4}. 

\section{Experimental relevance}
\label{Sec6}

Let us now investigate the experimental relevance of our results. First
of all, as mentioned also in Sec.~\ref{Sec1}, given the remarkable
progress in trapping and detecting single atoms, there is a general
tendency in the field of cold atomic gases towards the few-body limit
\cite{Wenz2013,Greene2017}, where interesting, finite-$N$
effects, are expected to show up. 

In order to confirm the derived results one could start with a
Bose-Einstein condensed cloud of atoms confined in either an annular,
or a toroidal potential, as in the experiments of Refs.\,\cite{Gupta2005,
Olson2007, Ryu2007, Sherlock2011, Ramanathan2011, Moulder2012,
Beattie2013, Ryu2014, Eckel2014, Navez2016}. Then, angular momentum
could be imparted to the system, as in Refs.\,\cite{Ryu2007, Sherlock2011,
Ramanathan2011, Moulder2012, Beattie2013, Ryu2014, Eckel2014, Navez2016}. 
Given the intimate connection between the solitary-wave and the ``yrast"
states \cite{Jackson2011}, giving angular momentum to the gas will, at
least for a sufficiently large number of atoms $N$, result in
solitary-wave state(s). According to our
study, if one reduces the number of particles in the system, the pure traveling-wave
solutions are no longer present. Instead, as the density disturbances travel around the ring, they also undergo decays,
as well as collapses and revivals. Alternatively, as we have also shown, one may utilize a stirring potential to set a few-body system rotating and obtain a similar behavior. The collapses and revivals are a very well-defined prediction of our study, which can be confirmed experimentally. The scaling of the corresponding characteristic timescales with the atom number $N$ may also be confirmed experimentally. 

Turning to the assumptions we have made, in experiments using toroidal traps, typically the atom number $N \sim 10^5$, the scattering length $a_s \sim 100$ \r{A}, the radius of the torus $R \sim 20$ $\mu$m and the characteristic transversal length $a_w \sim 5$ $\mu$m, resulting in values of $\gamma = 2 N a_s R/S \sim 10^3$, where $S = \pi(a_w/2)^2$ is the cross section of the torus. At first, the interaction strengths chosen here, $\gamma = 0.05$ and $0.2$, may strike as being far weaker than the value of  $\gamma$ typically encountered in many experiments. However, since $\gamma \propto N$, in the few-body regime our choices of $\gamma$ appear reasonable. We anticipate similar characteristic features for dynamical systems with values of $\gamma$ beyond those considered here.  By reducing the number of atoms $N$, also the chemical potential of the system is lowered. Such a reduction in the chemical potential would make excitations in the transverse direction less likely and thus strengthen the validity of our quasi one-dimensional treatment of the system.

Based on the two-state model, we may also estimate the periodicity $T_{GP} \approx 4\pi MR^2/\hbar$ of the initial travelling wave with $\ell=1/2$, see Eq.~(\ref{TGP}), as well as its revival time $T_A \approx NT_{GP}$. In particular, with $R \sim 20$ $\mu$m and $N \sim 10$, the considered time-scales become $T_{GP} \sim 7$ s and $T_{A} \sim 70$ s in the case of ${}^{87}$Rb. Obviously, both these time-scales can be reduced by considering systems of lighter (or fewer) atoms as well as ring potentials with shorter radii.
We therefore strongly believe that the results derived here is within reach of modern experimental techniques.

\section{Summary and conclusions}

\label{Sec7}

We  compared the temporal behavior of a Bose-Einstein condensate 
within the mean-field Gross-Pitaevskii equation with the full many-body 
description for finite $N$. 
While the time-dependent Gross-Pitaevskii equation is very 
successful in the case of a large particle number, its validity becomes 
questionable when $N$ becomes smaller, i.e., of order unity.

The time evolution of a single-product state representing a 
solitary-wave solution is particularly simple within the mean-field 
approximation. The single-particle density associated with such a state 
propagates around the ring without changing its shape. Time-evolved in 
agreement with the many-body Schr\"{o}dinger equation, however, the same 
kind of initial single-product state exhibits a more complex dynamical 
behavior. We found three characteristic timescales associated with 
different mechanisms. The first timescale is for a single revolution of 
the density peak around the ring, the behavior predicted by the 
mean-field approximation. Within the second timescale, the density peak 
of the solitary wave spreads and the distribution becomes more 
homogeneous. Within the third timescale, the system undergoes a single 
cycle in a periodic decay and revival of the initially inhomogeneous 
distribution. In the appropriate large-$N$ limit, these timescales show a 
clear hierarchy, being separated by a factor of $\sqrt{N}$. For 
increasing $N$, the full many-body dynamics thus approaches the mean-field 
results.

In addition, we have shown that the dynamics driven by an 
external potential $V$ stirring the few-body system exhibits the same behavior, with collapses and 
revivals of a solitary-wave state, after a sudden removal of $V$. In fact, 
the single-particle density of the initially stirred system is almost 
identical to that of an initial mean-field state with matching average 
energy and angular momentum.

Finally, with continued experimental progress towards trapping, manipulating and detecting fewer particles, we believe that the results of our study is of experimental interest.

\ack
We thank M. Magiropoulos, A. Roussou and J. Smyrnakis for useful discussions. 
This work was supported by the Swedish Research Council and NanoLund at Lund University.


\vskip 2em

\begin{thebibliography}{99}

\bibitem{Gupta2005} S. Gupta, K. W. Murch, K. L. Moore, T. P. Purdy, and D. M.
Stamper-Kurn, Phys. Rev. Lett. {\bf 95}, 143201 (2005).

\bibitem{Olson2007} Spencer E. Olson, Matthew L. Terraciano, Mark Bashkansky,
and Fredrik K. Fatemi, Phys. Rev. A {\bf 76}, 061404(R) (2007).

\bibitem{Ryu2007} C. Ryu, M. F. Andersen, P. Clad\'e, Vasant Natarajan,
K. Helmerson, and W. D. Phillips, Phys. Rev. Lett. {\bf 99}, 260401 (2007).

\bibitem{Sherlock2011} B. E. Sherlock, M. Gildemeister, E. Owen, E. Nugent, and
C. J. Foot, Phys. Rev. A {\bf 83}, 043408 (2011).

\bibitem{Ramanathan2011} A. Ramanathan, K. C. Wright, S. R. Muniz, M. Zelan,
W. T. Hill, C. J. Lobb, K. Helmerson, W. D. Phillips, and G. K. Campbell,
Phys. Rev. Lett. {\bf 106},  130401 (2011).

\bibitem{Moulder2012} Stuart Moulder, Scott Beattie, Robert P. Smith, Naaman
Tammuz, and Zoran Hadzibabic, Phys. Rev. A {\bf 86}, 013629 (2012).

\bibitem{Beattie2013} Scott Beattie, Stuart Moulder, Richard J. Fletcher, and
Zoran Hadzibabic, Phys. Rev. Lett. {\bf 110}, 025301 (2013).

\bibitem{Ryu2014} C. Ryu, K. C. Henderson and M. G. Boshier, New J. Phys.
{\bf 16}, 013046 (2014).

\bibitem{Eckel2014} Stephen Eckel, Jeffrey G. Lee, Fred Jendrzejewski,
Noel Murray, Charles W. Clark, Christopher J. Lobb,William D. Phillips,
Mark Edwards, and Gretchen K. Campbell, Nature (London) {\bf 506}, 200
(2014).

\bibitem{Navez2016} P. Navez, S. Pandey, H. Mas, K. Poulios, T. Fernholz, and
W. von Klitzing, New Journal of Phys. {\bf 18}, 075014 (2016).

\bibitem{Wenz2013} A. N. Wenz, G. Z\"urn, S. Murmann, I. Brouzos, T. Lompe, and S. Jochim,
Science {\bf 342}, 457 (2013). 

\bibitem{Jackson2001} A. D. Jackson, G. M. Kavoulakis, B. Mottelson, and S. M. Reimann,
Phys. Rev. Lett. {\bf 86}, 945 (2001)

\bibitem{Cremon2013} J. C. Cremon, G. M. Kavoulakis, B. R. Mottelson, 
and S. M. Reimann, Phys Rev A 87, 053615 (2013); 

\bibitem{Cremon2015} J. C. Cremon, A. D. Jackson, 
E. \" O. Karabulut, G. M. Kavoulakis, B. R. Mottelson, S. M. Reimann, Phys. Rev. A {\bf 91}, 
033623 (2015); 

\bibitem{Roussou2015} A. Roussou, G. D. Tsibidis, J. Smyrnakis, M. Magiropoulos, Nikolaos K. Efremidis, 
A. D. Jackson, G. M. Kavoulakis, Phys. Rev. A {\bf 91}, 023613 (2015). 

\bibitem{Scott1973} A. C. Scott, F. Y. F. Chu, and D. W. McLaughlin, Proc.
IEEE {\bf 61}, 1443 (1973).

\bibitem{Rajaraman1987} R. Rajaraman, {\it Solitons and Instantons}
(North-Holland, Amsterdam, 1987).

\bibitem{Zakharov1973} V. E. Zakharov and A. B. Shabat, Zh. Eksp. Teor. Fiz.
{\bf 64}, 1627 (1973) [Sov. Phys. JETP {\bf 37}, 823 (1973)].

\bibitem{Wadati1984} M. Wadati and M. Sakagami, J. Phys. Soc. Jpn. {\bf 53},
1933 (1984).

\bibitem{Wadati1985} M. Wadati, A. Kuniba, and T. Konishi, J. Phys. Soc. Jpn.
{\bf 54}, 1710 (1985).

\bibitem{Lai1989} Y. Lai and H. A. Haus, Phys. Rev. A {\bf 40}, 854 (1989).

\bibitem{Girardeau2000} M. D. Girardeau and E. M. Wright, Phys. Rev. Lett. {\bf 84},
5691 (2000).

\bibitem{Dziarmaga2002} Jacek Dziarmaga, Zbyszek P. Karkuszewski, and Krzysztof
Sacha, Phys. Rev. A {\bf 66}, 043615 (2002).

\bibitem{Berman2004} G.P. Berman, F. Borgonovi, F.M. Izrailev, and A. Smerzi, Phys. Rev. Lett. {\bf 92}, 030404 (2004).

\bibitem{Li2005} Weibin Li, Xiaotao Xie, Zhiming Zhan and Xiaoxue Yang, Phys,. Rev. A {\bf 72}, 043615 (2005).

\bibitem{Kanamoto2010} R. Kanamoto, L. D. Carr, and M. Ueda, Phys. Rev. A
{\bf 81}, 023625 (2010).

\bibitem{Zollner2011} S. Z\"ollner, G.M. Bruun, C.J. Pethick, and S.M. Reimann, Phys. Rev. Lett. {\bf 107}, 035301 (2011).

\bibitem{Sato2012} Jun Sato, Rina Kanamoto, Eriko Kaminishi, and Tetsuo
Deguchi, Phys. Rev. Lett. {\bf 108}, 110401 (2012).

\bibitem{Heimsoth2012} M. Heimsoth, C.E. Creffield, L.D. Carr and F. Sols, New J. Phys. {\bf 14}, 075023 (2012). 

\bibitem{Delande2014} Dominique Delande and Krzysztof Sacha, Phys. Rev. Lett.
{\bf 112}, 040402 (2014).

\bibitem{Syrwid2015} A. Syrwid and K. Sacha, e-print ArXiv:
1505.06586.

\bibitem{Syrwid2016} A. Syrwid, M. Brewczyk, M. Gajda, and
K. Sacha, e-print: ArXiv: 1605.08211.

\bibitem{Roussou2017} A. Roussou, J. Smyrnakis, M. Magiropoulos, Nikolaos K.
Efremidis, and G. M. Kavoulakis, Phys. Rev. A {\bf 95}, 033606 (2017).

\bibitem{Carr2000} L. D. Carr, C. W. Clark, and W. P. Reinhardt, Phys. Rev. A {\bf 62}, 063610 (2000).

\bibitem{Smyrnakis2010} J. Smyrnakis, M. Magiropoulos, G. M. Kavoulakis, and A. D. Jackson,
Phys. Rev. A {\bf 82}, 023604 (2010).

\bibitem{Jackson2011} A. D. Jackson, J. Smyrnakis, M. Magiropoulos, and G. M. Kavoulakis, 
Europh. Lett. {\bf 95} 30002, (2011).

\bibitem{Bloch1973} F. Bloch, Phys. Rev. A {\bf 7}, 2187 (1973).

\bibitem{Eberly1980} J. H. Eberly, N. B. Narozhny, and J. J. Sanchez-Mondragon,
Phys. Rev. Lett. {\bf 44}, 1323 (1980).

\bibitem{Park1986} T.~J. Park and J.~C. Light, J. Chem. Phys. {\bf 85}, 5870 (1986).

\bibitem{Lawson1967} J.~D. Lawson, {\it SIAM Journal on Numerical Analysis}, Vol.~4, p.~372 (1967).
  
\bibitem{Burger1999}  S. Burger, K. Bongs, S. Dettmer,
W. Ertmer, K. Sengstock, A. Sanpera, G. V. Shlyapnikov, and
M. Lewenstein, Phys. Rev. Lett. {\bf 83}, 5198 (1999).

\bibitem{Andersen2006} M. F. Andersen,
C. Ryu, Pierre Cladé, Vasant Natarajan, A. Vaziri, K. Helmerson,
and W. D. Phillips, Phys. Rev. Lett. {\bf 97}, 170406 (2006).

\bibitem{Greene2017} C. H. Greene, P. Giannakeas, and J. Perez-Rios
Rev. Mod. Phys. {\bf 89}, 035006 (2017)
  
\end{thebibliography}
\end{document}